\documentclass[a4paper,11pt]{article}

\usepackage{jheppub}

\usepackage[T1]{fontenc} 
\usepackage{subcaption}
\usepackage{verbatim}
\usepackage{mathrsfs}
\usepackage{physics}
\usepackage{amsthm}
\usepackage{mathtools}
\usepackage{enumerate}
\usepackage{graphicx}

\renewcommand{\tilde}{\widetilde}
\renewcommand{\bar}{\overline}

\renewcommand{\H}{\mathcal{H}}

\title{Notes on the Factorisation of the Hilbert Space for Two-Sided Black Holes in Higher Dimensions}
\author{Pan Li $^{1,2}$}
\affiliation{$^1$Institute of High Energy
Physics, Chinese Academy of Sciences, Beijing 100049, China\\ $^2$
School of Physics, University of Chinese Academy of Sciences,
Beijing 100049, China }
\emailAdd{lipan@ihep.ac.cn} 
\abstract{In this paper, we investigate the Hilbert space factorisation problem of two-sided black holes in high dimensions. We demonstrate that the Hilbert space of two-sided black holes can be factorized into the tensor product of two one-sided bulk Hilbert spaces when the effect of non-perturbative replica wormholes is taken into account. We further interpret the one-sided bulk Hilbert space as the Hilbert space of a one-sided black hole. Therefore, since the Hilbert space of a two-sided black hole can be obtained from the tensor product of two single-sided black hole Hilbert spaces, we consider this as an embodiment of the ER=EPR conjecture, and we show when the entanglement between the two single-sided black holes is sufficiently strong, the (Lorentzian) geometry of a two-sided black hole will emerge.}

\makeatletter
\gdef\@fpheader{\vspace{-1.5em}}
\makeatother
\begin{document}
\maketitle

\section{Introduction}

The semi-classical description of gravity would lead to some puzzles, which are generally believed not to appear in a complete and well-defined non-perturbative theory of quantum gravity. Among the puzzles exemplified by the black hole information problem \cite{Hawking:1976ra}, one of the most notable issues is that semi-classical gravity appears to encompass an excessive number of quantum black hole microstates, far exceeding the Bekenstein-Hawking entropy, which is given by ${A}/{4G}$. For a long time, it has been believed that resolving such an issue might require a UV-complete quantum theory of gravity. However, by recent advances in lower-dimensional quantum gravity theories \cite{Saad:2019lba} and the resulting replica wormhole formulation \cite{Penington:2019kki}\cite{Almheiri:2019qdq}, significant breakthroughs have been achieved in the black hole information problem. The replica wormhole formalism leads to the result that perturbative orthogonal black hole microstates are not genuinely orthogonal. Instead, the inner product of perturbative orthogonal states will receive  non-perturbative  corrections from the replica wormhole with order $e^{-S}$, which means
\begin{align}
   & \bra{\psi_i}\ket{\psi_j}_{pert}=\delta_{ij}, \\
   &|\bra{\psi_i}\ket{\psi_j}|^2_{non-pert}=\delta_{ij}+e^{-S}.
\end{align}
where $S$ is the entropy of the black hole. When such non-perturbative corrections are considered, the naive infinite number of perturbative black hole microstates becomes finite and matches ${A}/{4G}$.

As shown in \cite{Boruch:2024kvv}, the wormhole corrections could also be used to address another controversy between semi-classical gravity and non-perturbative quantum gravity, namely the ``Hilbert space factorisation problem''.  This issue becomes particularly evident in the context of two-sided AdS black holes. 

From a semi-classical perspective, since the fixed smooth geometry of the two-sided black hole connects two asymptotic boundaries, the perturbative states on the black hole background may just be described by the quantum field theory in curved spacetime (QFTCS). Actually, for a ``semi-classical  gravitational state", we could think of it as  the ``tensor product'' of gravitational states and matter states, where  we just simply take the classical geometry of the background as the gravitational state, and the matter state is the normal QFTCS state, where the total Hilbert space is 
 \begin{equation}
     \H_{pert}=\H_{grav}\otimes \H_{matter}.
 \end{equation}
 Therefore, it will be easy to construct an infinite set of perturbative orthogonal states (i.e. infinite kinds of flavor scalar fields) just as in the normal quantum field theory. If we construct excited states of quantum fields inside the black hole, it seems that we naively obtain an infinite number of black hole microstates at perturbative level \cite{Balasubramanian:2022gmo}. However, this contradicts the requirement from black hole thermodynamics that the black hole has a finite entropy ${A}/{4G}$.

From the non-perturbative perspective, as pointed out by \cite{Maldacena:2001kr}, in AdS/CFT, the two-sided AdS black hole state (Hartle-Hawking state) is dual to the thermal field double (TFD) state, which is an entangled state of two non-interacting conformal field theories (CFTs). Thus, the TFD state belongs to the tensor product of the Hilbert spaces of two CFTs. Also, the excited states on the two-sided black hole background should all belong to the tensor product of two CFT Hilbert spaces
\begin{equation}
    \H_{\text{non-pert}
    }=\H_L^{CFT}\otimes \H_R^{CFT}.
\end{equation}

We have demonstrated that perturbative and non-perturbative Hilbert spaces of two-sided AdS black holes are completely distinct. As we mentioned above, wormhole effects will cure the naive infinite dimensional $\H_{pert}$. Moreover, in this paper, following the idea of  \cite{Boruch:2024kvv} in two dimensions JT gravity, we will show that in higher dimensional spacetime, wormhole effects will also make the naive perturbative Hilbert space become factorisation, moreover, it will factorize into the tensor product of two single-sided black hole Hilbert space, which reflects the idea of ER=EPR conjecture.

The structure of this paper is as follows: in section \ref{sec:thinshell}, we will show how to construct the infinite number of perturbative states via the excitations in the interior of a black hole \cite{Balasubramanian:2022gmo}. In section \ref{Factorisation of Hilbert space}, we will prove the factorisation of the Hilbert space in higher dimensional space time with the use of the resolvent formalism. In section \ref{sec:From factorization to ER=EPR}, we will discuss the relationship between the Hilbert space factorisation and ER=EPR conjecture.

Note added:
As our work was nearing completion, we learned of \cite{Balasubramanian:2024yxk}, which also examines the Hilbert space factorisation problem in the context of higher-dimensional two-sided black hole backgrounds.

\section{Thin shell states as black hole microstates}
\label{sec:thinshell}

In this section, we will briefly review how to construct the microstates of black holes using thin shell states; see \cite{Chandra:2022fwi}(single-sided black holes) and \cite{Balasubramanian:2022gmo}(two-sided black holes) for details. From a statistical mechanics perspective, a macroscopic state corresponds to multiple microstates. This means that for a given equilibrium system with specified macroscopic parameters (such as \( N, V, T \)), there can be many microstates that share the same macroscopic parameters, but the specific micro-level information differs. There is ample evidence indicating that black holes are indeed thermodynamic systems \cite{Wald:1999vt}, allowing us to study their statistical mechanical properties. The macroscopic parameters of equilibrium black holes—such as ADM mass, charge, and angular momentum—are entirely determined by the geometry outside the event horizon. This implies that black holes with the same external geometry belong to the same macroscopic state. However, identifying the microstates of a general black hole (such as a Schwarzschild black hole) remains an open question.

Recently, it was shown that the microstates of general black holes can be constructed using thin shell excitations within the black hole interior \cite{Chandra:2022fwi}\cite{Balasubramanian:2022gmo}, extending the previous two-dimensional case \cite{Kourkoulou:2017zaj}\cite{Goel:2018ubv}. Thin shell black holes refer to a class of black holes that have the same geometry outside the event horizon (indicating they correspond to the same macroscopic state), while inside, they possess different thin shell matter (which can be regarded as different microstates). In the classical limit, the solutions inside and outside the thin shell are connected by Israel junction conditions.

In the quantum mechanical sense, black holes with different interiors will be described by different quantum states, where the excitations in the black hole interior arise from operator insertions during the Euclidean path integral in the state preparation process. In $\text{AdS}_3/\text{CFT}_2$, the relationship between operator insertions in the conformal field theory and thin shell black holes has been investigated in detail \cite{Chandra:2024vhm}. Following we will construct thin shell states as two-sided and one-sided black hole microstates separately. 

 \subsection{ The geometry of thin shell black holes}
 \label{sec:geometry-of-bh}
 \subsubsection*{The case of two-sided black holes}
In this section, we will describe how to construct the geometry of two-sided black holes with a thin shell; further details can be found in \cite{Balasubramanian:2022gmo}.

\begin{figure}[ht]
    \centering
    \begin{subfigure}{0.4\textwidth}
        \centering
        \includegraphics[width=\linewidth]{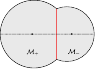} 
        \caption{Euclidean thin shell solution}
        \label{fig:euclidean-thin-shell-solution}
    \end{subfigure}
    \hfill
    \begin{subfigure}{0.4\textwidth}
        \centering
        \includegraphics[width=\linewidth]{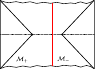} 
        \caption{Lorentzian thin shell solution }
        \label{fig:Lorentzian thin shell solution}
    \end{subfigure}
    \caption{the thin shell solution consist of $\mathcal{M}=\mathcal{M_+}\cup \text{shell} \cup \mathcal{M_-}$. Following the standard approach, we prepare the quantum state using the Euclidean path integral and then analytically continue it to Lorentzian spacetime on the \(t = 0\) slice (dashed line).In the Lorentzian picture, the thin shell propagates from singularity to singularity.}
    \label{fig:E and L thin shell solutions}
\end{figure}

We will consider the following two-sided black holes: they can be thought of as the extension of eternal Schwarzschild\footnote{Schwarzschild implies AdS-Schwarzschild.} black hole: outside the horizon, they are simply the eternal Schwarzschild black hole solution, while in the interior, the presence of the thin shell enlarges the interior of the black hole and causes the horizons on the left and right sides to no longer coincide, see Fig \ref{fig:E and L thin shell solutions}. The thin shell is always located behind the horizon if the mass of the shell is large enough, and we will only consider this case.

We construct the Euclidean solution of the two-sided thin-shell black hole as follows: the left side ($\mathcal{M_+} $) and right side ($\mathcal{M_-} $) of the thin shell are both eternal Schwarzschild black hole solutions, but they can generally have different masses due to the presence of the thin shell. Locally, they are described by  Schwarzschild metric

\begin{equation}
 d s_{ \pm}^2=f_{ \pm}\left(r\right) d \tau^2+\frac{d r^2}{f_{ \pm}\left(r\right)}+r^2 d \Omega_{d-1}^2
\end{equation}
with 
\begin{equation}
    f_{\pm}(r)=1+r^2-\frac{M_{ \pm}}{r^{d-2}}.
\end{equation}

At the interface formed by the spherical thin shell, we apply the Israel junction conditions to connect the solutions on the left and right sides: 
\begin{equation}
    \Delta K_{a b}-\Delta K h_{a b}=-S_{a b},
\end{equation}
where $ \Delta K_{a b}$ represents the difference in extrinsic curvature on two sides of the shell and  $S_{ab}$  is the surface stress tensor for the shell. For simplicity, we assume the matter is a pressureless perfect fluid, thus the form of $S_{ab}$ is known, and then the trajectory of the shell is determined by the Israel junction condition. 

As a summary, consider a two-sided black hole which is described by Schwarzschild geometry outside the horizon and a spherically symmetric thin shell formed by a pressureless ideal fluid inside the horizon, then the entire solution is fully determined by three parameters: the ADM mass of the left black hole $M_+$, the ADM mass of the right black hole $M_-$, and the local mass of the thin shell\footnote{We refer to it as the local mass because this mass does not directly manifest in the ADM mass. In Lorentzian geometry, the thin shell propagates from the past singularity to the future singularity. Therefore, we can treat this mass as a parameter in the local effective description.}.

\subsubsection*{The case of single-sided black hole}
The philosophy behind constructing two-sided and one-sided black hole thin shell solutions is the same. The difference lies in that the exterior of the one-sided black hole still follows the Schwarzschild geometry, but the interior is now described by AdS vacuum \cite{Chandra:2022fwi}. The trajectory of the thin shell is still determined by the Israel junction condition. This solution is analogous to the Oppenheimer-Snyder solution, but with the distinction that it is time-reversal symmetric. 
\begin{figure}[ht]
    \centering
    \begin{subfigure}{0.25\textwidth}
        \centering
        \includegraphics[width=\linewidth]{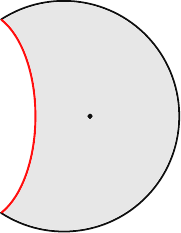} 
        \caption{Euclidean thin shell solution}
        \label{fig:SINGLE euclidean-thin-shell-solution}
    \end{subfigure}
    \hspace{0.2\textwidth}
    \begin{subfigure}{0.3\textwidth}
        \centering
        \includegraphics[width=\linewidth]{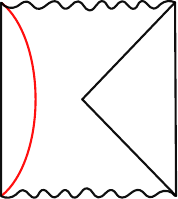} 
        \caption{Lorentzian thin shell solution }
        \label{fig:SINGLE Lorentzian thin shell solution}
    \end{subfigure}
    \caption{The geometry inside the thin shell is an AdS vacuum, while the geometry outside the thin shell is described by a AdS-Schwarzschild solution. Similar to the two-sided case, in Lorentzian geometry, the matter propagates from one singularity to the other.}
    \label{fig:E and L single sided thin shell solutions}
\end{figure}

 \subsection{The quantum state interpretation of thin shell black holes}
 \label{sec:thinshell_state}
 \subsubsection*{two-sided black holes}
 
Now, we want to describe the corresponding quantum states of the thin shell black holes. Firstly, we will review a famous example: as pointed out by \cite{Maldacena:2001kr}, in AdS/CFT, the (unnormalized) thermal-field double state
\begin{equation}
    \ket{\Psi(\beta)}= \sum_{n} e^{-\frac{1}{2}\beta E_n}\ket{E_n}_L\ket{E_n}^*_R,
\end{equation}
will approximately correspond to the two-sided eternal Schwarzschild black hole. The correspondence can be understood as follows: the inner product of $\ket{\Psi(\beta)}$ can be represented as a Euclidean path integral, which is the trace of the thermal density matrix
\begin{equation}
Z(\beta)=\bra{\Psi(\beta)}\ket{\Psi(\beta)}=
    \Tr(e^{-\beta H})=\oint_{\phi(0)=\phi(\beta)} D\phi e^{-I_E(\phi)} \approx e^{-I(\beta)}.
\end{equation}
Using AdS/CFT, the trace can also be understood as that of the Euclidean gravitational path integral (EGPI). In the computation of the EGPI, We can use the saddle point solution as an approximation of the full path integral. Here, $I(\beta)$ denotes the on-shell action of the Euclidean black hole solution with inverse temperature $\beta$, therefore, we conclude that the TFD state corresponds to the two-sided eternal black hole. When the temperature is above the Hawking-Page transition point, the dominant saddle solution is the Euclidean black hole solution.

We can represent the path integral using a disk diagram
\begin{equation}
    \bra{\Psi(\beta)}\ket{\Psi(\beta)} \approx e^{-I(\beta)}=\vcenter{\hbox{\includegraphics[width=0.2\textwidth]{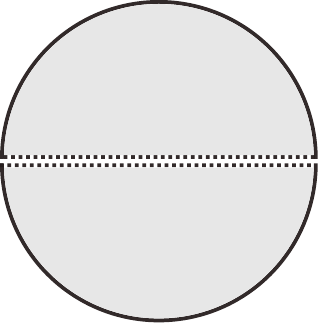}}} .
\end{equation}

Similarly, the thin shell black hole could also be achieved from the saddle point approximation solution of the EGPI \cite{Balasubramanian:2022gmo}, which is a higher-dimensional extension of the two-dimensional ``partially entangled thermal states'' \cite{Goel:2018ubv}. In the dual CFT theory, the thin shell in the interior of the black hole corresponds to heavy operator insertions. More specifically, the conformal dimension of the operator $\hat{O_i}$ (composed of a large number of single-trace operators) corresponds to the mass $m_i$ of the thin shell. For a good semi-classical approximation, we also require the operators to be very non-local and uniformly distributed on the spherical surface in the CFT. Specifically, the thin shell states have the form of
\begin{equation}
\ket{\Psi_i(\tilde{\beta}_L,\tilde{\beta}_R)} =  \sum_{m,n} e^{-\frac{1}{2}\tilde{\beta}_LE_m-\frac{1}{2}\tilde{\beta}_R E_n}(O_i)_{mn}\ket{E_m}_L\ket{E_n}_R
       =\ket{\rho_{\tilde{\beta}_L/2}\, O_i \, \rho_{\tilde{\beta}_R/2}}.
\end{equation}
Note that due to the existence of shell, the left and right black holes could have different temperatures, which has already been seen from the thin shell black hole geometry, and here we use $\tilde{\beta}$ because the physical $\beta$  generally is not $\tilde{\beta}$ (this will be explained in detail later).  The inner product and the diagram representation are
\begin{align}
\bra{\Psi_i(\tilde{\beta}_L,\tilde{\beta}_R)} \ket{\Psi_i(\tilde{\beta}_L,\tilde{\beta}_R)}&=\Tr\left(O^\dagger_i\, \rho_{\tilde{\beta}_L}\, O_i \, \rho_{\tilde{\beta}_R}  \right) \\ &\approx e^{-I_i(\tilde{\beta}_L,\tilde{\beta}_R)}=\vcenter{\hbox{\includegraphics[width=0.13
\textwidth]{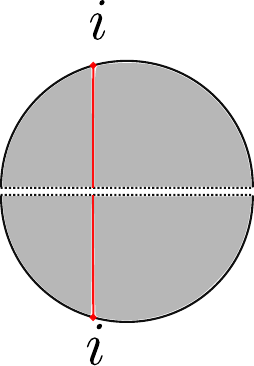}}} \, ,
\end{align}
where the saddle point action $I_i(\tilde{\beta}_L,\tilde{\beta}_R)$ contains both the on-shell gravitational action and the contribution from thin shells, which can be computed from Eq. (\ref{eq:shell-action}). In the following, $\tilde{\beta}_{L / R}$ will be unimportant for us, so for simplicity we just use 
$\ket{\Psi_i}$  and $I_i$. We may restore the complete notation when necessary.

\subsubsection*{single-sided black holes}
In this section, we aim to briefly describe the quantum states of single-sided thin shell black holes\footnote{see also \cite{Geng:2024jmm} for the construction of black hole states with the end of world branes.} \cite{Chandra:2022fwi}\cite{Anous:2016kss}. 
Similar to the two-sided case, in the CFT description, firstly we can form the shell state via acting uniformly smeared operators on the vacuum, which results in an unnormalized state $\ket{S}$. Then, by applying Euclidean time evolution on $\ket{S}$, we obtain the state 
\begin{equation}
    \ket{\psi(\tilde{\beta})}=e^{-\frac{\tilde{\beta}}{2} H}\ket{S}=e^{-\frac{\tilde{\beta}}{2} H}\hat{O}\ket{0}.
\end{equation}
This state will correspond to a single-sided thin shell black hole; see \cite{Anous:2016kss} for details. When we compute the inner product of this state using the path integral, the saddle point approximation of the path integral will give the geometry of the single-sided thin shell black hole, as shown in Fig \ref{fig:SINGLE euclidean-thin-shell-solution}. Therefore, one may construct a set of thin shell states for single-sided black holes with different masses of the shells by applying different operators $\hat{O}_i$.

\subsubsection*{large shell mass limits}
\label{large-mass-limit}
In the previous two sections, we have presented the construction of the thin shell black hole geometry and the corresponding quantum states. We aim to further establish the parameter correspondence between the geometry and the quantum states. Taking the two-sided black hole as an example, we have already mentioned that the black hole solution is entirely determined by the parameters of the ADM mass and the thin shell mass $m$, with the temperature being a function of the ADM mass. In contrast, the quantum states involve parameters $\tilde{\beta}_L$ and $\tilde{\beta}_R$. Due to the insertion of operators, these do not correspond to the physical temperature. In general, $\tilde{\beta}_{L/R}$ is a function of the physical temperature parameter $\beta_{L/R}$ (ADM mass) and the shell mass $m$. However, as shown by \cite{Balasubramanian:2022gmo}, 
in the large $m$ limit, i.e., when $m$ is much greater than the ADM mass, the corresponding relationship will be greatly simplified, which are \footnote{Note that, in the absence of operator insertions, \( \tilde{\beta}_L \) and \( \tilde{\beta}_R \) individually have no direct relation to temperature; they are simply two parameters, and the temperatures on both sides of the black hole are equal, with the inverse temperature $\beta$ being \((\tilde{\beta}_L+\tilde{\beta}_R )\),  } 
\begin{align}
    \tilde{\beta}_L&\approx\beta_L, \\
    \tilde{\beta}_R&\approx\beta_R.
\end{align}
After the insertion of a thin shell with a large mass, $\tilde{\beta}_{L/R}$ will decouple with $m$. Thus, we may have infinitely many states with different values of $m_i$, but the same values of $\beta_{L/R}$. And in the large $m$ limit, the inner product of the states will be very simple as we will see below. The same simplification also appears in the calculations for the single-sided black hole. Thus, we will only consider the large $m$ limit from now on, moreover, when discussing different masses \( m_i \) and \( m_j \), we require their difference to be sufficiently large to ensure they appear distinct at the semi-classical level.

\subsection{Non-perturbative overlap of thin shell states and replica wormholes}
\label{sec:overlaps}

In this section, we will use the EGPI to calculate the overlap between different thin-shell states. Moreover, we will explore how non-perturbative saddles come into play.

Firstly, we will compute the single inner product $\bra{\Psi_{i_1}}\ket{\Psi_{i_2}}$. At the perturbative level where $G\rightarrow0$, different thin shell states are naively orthogonal due to the different local thin shell mass $m$. We have
\begin{equation}
\label{eq:perturbative-inner-product}
\bra{\Psi_{i_1}}\ket{\Psi_{i_2}}=\vcenter{\hbox{\includegraphics[width=0.15\textwidth]{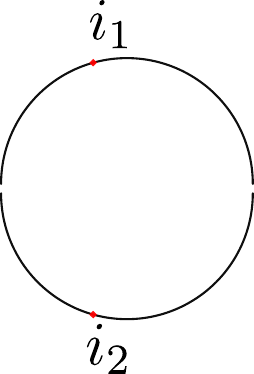}}} \approx\vcenter{\hbox{\includegraphics[width=0.15\textwidth]{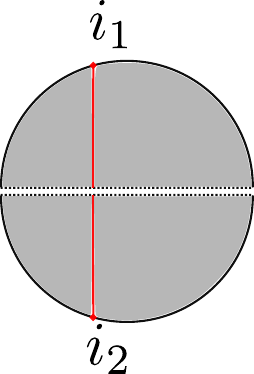}}}=Z_1 \delta_{i_1 i_2}. 
\end{equation}
The delta function means that the classical solutions only exist for $m_{i_1}=m_{i_2}$, which can be understood as the fact that a particle with mass \( m \) does not change its mass during its propagation. Here $Z_1$ is the saddle point approximation of the partition function of two-sided thin shell black holes. In the saddle point approximation, it can be obtained from the action
\begin{equation}
\label{eq:shell-action}
    I=-\frac{1}{16 \pi G} \int_{\text {bulk }} d^{d+1} x \sqrt{g}(R+\frac{d(d-1)}{l^2} )+\frac{1}{8 \pi G}\int_{\text {bdy }} d^{d} y \sqrt{h} K -m\int_{\text{shell}}dl,
\end{equation}
where $m$ is the mass of the shell. By substituting the black hole solutions obtained in \ref{sec:geometry-of-bh} into the action, we can calculate the on-shell action. Some counterterms are needed to get a finite action \cite{Balasubramanian:2022gmo}\cite{Climent:2024trz}, While for our purposes, we do not need to delve into the details of computing the action, as in the large shell mass limit, action of the thin-shell will decouple from the gravitational action. The action in the large shell mass limit \( m \gg M_{\pm} \) is very simple, which leads to the result that the unnormalized inner product of the thin shell state is just the product of the gravitational part and the matter part
\begin{equation}
\bra{\Psi_{i}}\ket{\Psi_{i}}\approx Z_1=Z_1(\beta_L)Z_1(\beta_R)\,e^{-2 m_i \log R_i^{*}},
\end{equation}
where $R_i^{*}$ is the minimal radius of the shell trajectory, and in the large \( m \) limit, \( R_i^* \sim \sqrt[d-1]{Glm_i} \). In the following sections, since the shell mass term is not actually important for our study (as it will be canceled out in the normalization), we will omit the shell mass term when calculating the inner product and consider only the gravitational action. That is, we will denote the inner product in the form of 
\begin{equation}
\bra{\Psi_{i}}\ket{\Psi_{i}}\approx Z_1=Z_1(\beta_L)Z_1(\beta_R).
\end{equation}

Non-trivial saddles arise when we compute the product of multiple inner products. In the manner of Euclidean gravitational path integral, 
each $\bra{\Psi_i}$ or $\ket{\Psi_j}$ defines a gravitational boundary condition. The principles of Euclidean gravitational path integrals tell us that, when computing the product of $n$ inner products, we should sum over all the possible topology which respect the $n$ boundary conditions, and perform the path integral over that manifolds. In the semiclassical limit, we consider only the contributions from the saddle point solutions. \footnote {In the case of JT gravity, the authors of \cite{Saad:2019lba} elegantly demonstrate the computation of the full Euclidean gravitational path integral by summing over different topologies, and establishing a profound connection between these topological sums and random matrix theory. While, performing a complete and reliable path integral for higher-dimensional gravity remains profoundly unclear, both conceptually and technically. Here we just consider two saddle point solutions: the disk saddle (disconnected) and the pinwheel saddle (connected). We expect that the above two saddle point solutions would provide the major contribution, at least for the problem we are considering. However, confirming this major contribution is crucial to our conclusions and presents a question worthy of further investigation. }

Therefore, according to the rules of EGPI, the norm square of the inner product is 
\begin{align}
\label{eq:phi square}
\bra{\Psi_{i_1}}\ket{\Psi_{i_2}}\bra{\Psi_{i_2}}\ket{\Psi_{i_1}}&=\vcenter{\hbox{\includegraphics[width=0.3\textwidth]{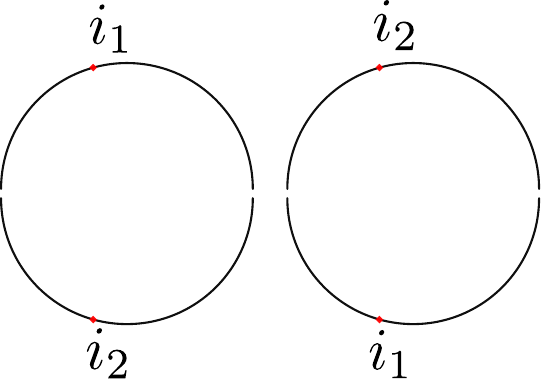}}}  \\
&\approx\vcenter{\hbox{\includegraphics[width=0.3\textwidth]{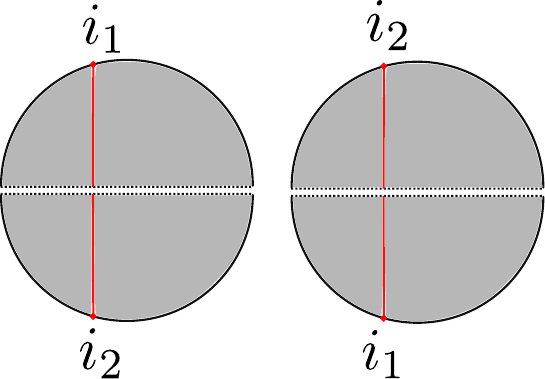}}}+\vcenter{\hbox{\includegraphics[width=0.15\textwidth]{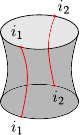}}} \\
&= Z_1^2 \delta_{i_1 i_2}+Z_2
\end{align}

In the case with four boundaries, the saddle point approximation gives us two possible geometries: the first is just like the two boundary case, only the same mass shell ($i_1=i_2$) in the inner product can be non-zero; while there is another solution with respect to the boundary condition coming from the replica geometry:  even $i_1\ne i_2$, in the replica geometry, the index $i_1$ can propagate to the other $i_1$ living on the boundary of its replica partner. It is not difficult to obtain that for the fully connected saddle solution $Z_n$
\begin{equation}
\bra{\Psi_{i_1}}\ket{\Psi_{i_2}}\bra{\Psi_{i_2}}\dots\bra{\Psi_{i_n}}\ket{\Psi_{i_1}}_c\approx Z_n=\vcenter{\hbox{\includegraphics[width=0.25\textwidth]{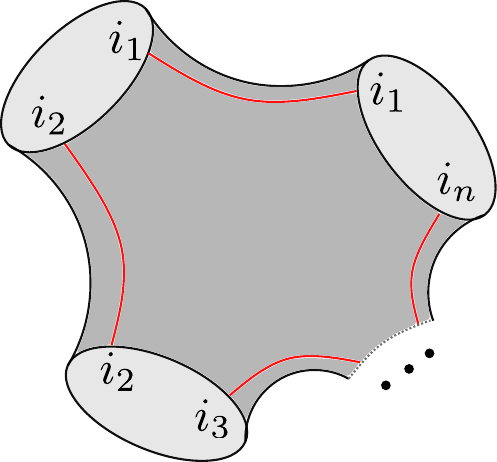}}}.
\end{equation}

Note that there are also some block-connected geometries, for example:
\begin{align}
&\bra{\Psi_{i_1}}\ket{\Psi_{i_2}}\bra{\Psi_{i_2}}\dots\bra{\Psi_{i_{n-3}}}\ket{\Psi_{i_{n-2}}}_c \,\cdot \,\bra{\Psi_{i_{n-2}}}\ket{\Psi_{i_{n-1}}}\bra{\Psi_{i_{n-1}}}\ket{\Psi_{i_{n}}}\bra{\Psi_{i_{n}}}\ket{\Psi_{i_{1}}}_c \\ &\approx \delta_{i_1 i_{n-2}} Z_{n-3} \cdot Z_3=\vcenter{\hbox{\includegraphics[width=0.4\textwidth]{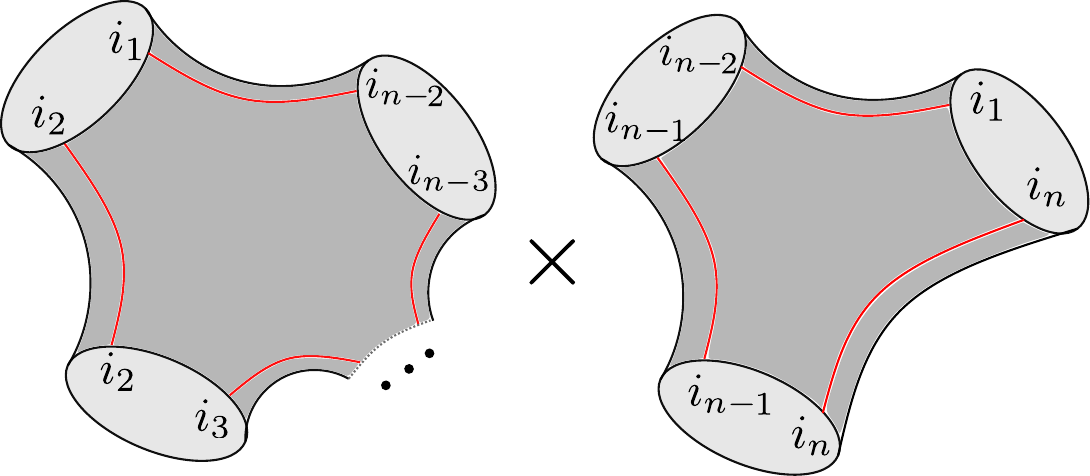}}}.
\end{align}

For a single-sided black hole state, using the Euclidean gravitational path integral, a similar result \cite{Chandra:2022fwi} could be obtained. We denote the fully connected saddle solution as
\begin{equation}
\bra{\psi_{i_1}}\ket{\psi_{i_2}}\bra{\psi_{i_2}}\dots\bra{\psi_{i_n}}\ket{\psi_{i_1}}_c\approx \tilde{Z}_n.
\end{equation}

It is well known that replica wormholes lead to the factorization puzzle of amplitudes, for example,
\begin{equation}
(\bra{\Psi_{i_1}}\ket{\Psi_{i_2}}\bra{\Psi_{i_2}}\ket{\Psi_{i_1}}) \ne (\bra{\Psi_{i_1}}\ket{\Psi_{i_2}}) \cdot (\bra{\Psi_{i_2}}\ket{\Psi_{i_1}})
\end{equation}
 
In JT gravity, this is understood as a characteristic feature of ensemble theories, as seen in dualities like RMT/JT \cite{Saad:2019lba}  or SYK/JT \cite{Maldacena:2016upp}. In contrast, for higher-dimensional, such as the $\text{AdS}_5/\text{CFT}_4$ duality, we know that it is not an ensemble theory. In this context, we may interpret the emergence of replica wormholes as a form of coarse-graining of a single theory. we adopt the perspectives of \cite{Sasieta:2022ksu} and \cite{Balasubramanian:2022gmo}: The semiclassical path integral calculates a coarse-grained average over the microstates that align with the macroscopic semiclassical description given by the saddle point. Therefore, the inner products we compute above should be understood in the sense of averaging or coarse-graining. 

\subsection{The Hilbert space construction from the thin shell states}
Following \cite{Boruch:2024kvv}, we can construct the Hilbert space via the different operator $\hat{O_i}$ acting on the reference state:
For a two-sided black hole, we can define
\begin{equation}
    \H_{\text{bulk(D)}}=\text{Span}\,\,\{{\ket{\Psi_i}}=\ket{\rho_{\beta_{L/2}}\, O_i \, \rho_{\beta_{R/2}}}|\,i=1, 2, \dots D\}.
\end{equation}
While for single-sided black holes, 
we can define
\begin{equation}
    \H_{L(d_L)}=\text{Span}\,\,\{{\ket{\psi_i}}=e^{-\frac{\beta H}{2}} O_i\ket{0}|\,i=1, 2, \dots d_L\}.
\end{equation}
As we know from Sec \ref{sec:thinshell_state}, the states in this Hilbert space are the ``microscopic states'' of black holes that possess the same geometry outside the horizon. Of course, it is easy to see that the dimension of $\H_{\text{bulk(D)}}$ and $ \H_{L(d_L)}$ depends on \( D \) and $d_L$. From the perspective of perturbation theory, neglecting the effects of wormholes, we can derive from Eq.(\ref{eq:perturbative-inner-product}) that all the thin shell states are naively orthogonal. However, as we will see above, the wormhole effects will introduce (very small) non-orthogonal overlaps among states that are orthogonal in the perturbative sense. Moreover, when \( D \) and \( d_L \) exceed their non-perturbative Hilbert space dimensions,  that is the exponent of black hole entropy. This seemingly small wormhole effect will correct the perturbative Hilbert space dimension to the non-perturbative Hilbert space dimension.

\section{Factorisation of Hilbert space}
\label{Factorisation of Hilbert space}
\subsection{Factorisation of trace}
Now we have all the essential tools to prove the factorisation of the bulk Hilbert space in higher dimensional spacetime. We want to prove such a factorisation relationship
\begin{equation}
\label{eq:factorisation}
    \Tr_{\mathcal{H}_\text{bulk}}(k_L k_R)=\Tr_{\mathcal{H}_L}(k_L)\Tr_{\mathcal{H}_R}(k_R),
\end{equation}
where $\H_{\text{bulk}}$ represents the non-perturbative bulk Hilbert space of the two-sided black hole, while $\H_L$ and $\H_R$ are the corresponding single-sided bulk Hilbert spaces, with  $k_L$ and $k_R$ being any function of the Hamiltonian $H_L$ and $H_R$ defined on single-sided Hilbert space $\H_L$ and $\H_R$. We will consider the micro-canonical ensemble for simplicity, and take into account the most general case, where the geometry corresponding to $\H_{\text{bulk}}$ is a two-sided black hole with different ADM masses for the left and right black holes. Then in a non-perturbative sense, the dimension of the Hilbert space for the two-sided black hole is given by $e^{S_L+S_R}$, obtained by multiplying the dimensions of the Hilbert spaces associated with the left and right side black holes, where $S_L$ and $S_R$ are the Bekenstein-Hawking entropy of the left and right side black holes.

\begin{figure}[t] 
    \centering
\includegraphics[width=1\textwidth]{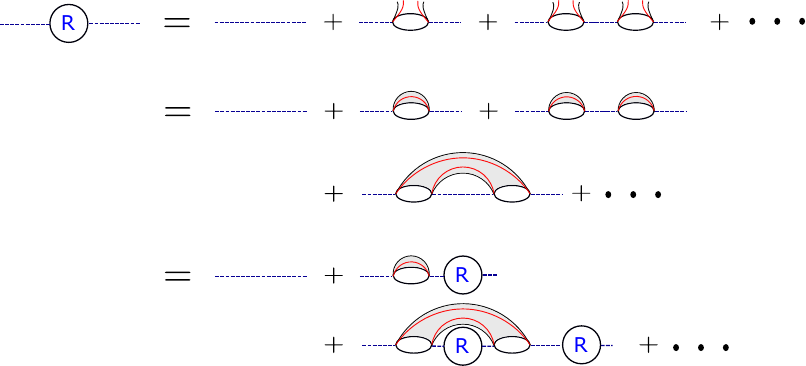} 
    \caption{resolvent} 
    \label{fig:resolvent} 
\end{figure}

To prove the factorisation of the trace in the bulk Hilbert space for a two-sided black hole, we should first figure out a way to represent the basis of the bulk Hilbert space. As demonstrated in \cite{Balasubramanian:2022gmo}, when the number of perturbative orthogonal states \( D \ge e^{S_L + S_R} \), the Hilbert space $\H_{\text{bulk(D)}}$ spanned by the thin shell states will equal the non-perturbative Hilbert space $\H_{\text{bulk}}$. Therefore, we can use the thin shell state as a basis for the two-sided black hole Hilbert space, that is, we need to prove the following equation:
\begin{equation}
    \Tr_{\mathcal{H}_\text{bulk(D)}}(k_L k_R)=\Tr_{\mathcal{H}_L}(k_L)\Tr_{\mathcal{H}_R}(k_R),
\end{equation}
under the condition $ D \ge e^{S_L + S_R}$.
Note the basis vectors (the thin shell state) in $\H_{\text{bulk(D)}}$ are non-orthogonal due to the non-perturbative wormhole effects, we should apply the trace formula for non-orthogonal bases:
\begin{equation}
\label{eq:(3.5)}
    \Tr_{\mathcal{H}_\text{bulk(D)}}(k_L k_R)=(M^{-1})_{ji}\bra{\Psi_i}k_Lk_R\ket{\Psi_j},
\end{equation}
where $\hat{M}$ is the Gram matrix
\begin{equation}
    M_{ij}=\bra{\Psi_j}\ket{\Psi_i}.
\end{equation}
We adopt the method in \cite{Boruch:2024kvv}  to first compute  $M^n$  for positive integer values of $n$, and then analytically continue $n$ to $-1$. 
we define the resolvent matrix $\hat{R}(\lambda)$ 

\begin{equation}
\label{eq:3.6}
    R_{ij}(\lambda)=(\frac{1}{\lambda-M})_{ij}=\frac{1}{\lambda}\sum_{n=0}^{\infty}\frac{(M^n)_{ij}}{\lambda^n},
\end{equation}
to solve $M^n$. By virtue of the residue theorem, when the integral curve of \( \lambda \) is a very large circle, such that all poles of $\hat{R}(\lambda)$ are within the circle. we may obtain 
\begin{equation}
    (M^n)_{ij}=\frac{1}{2 \pi i}\oint d\lambda \, \lambda^n R_{ij}(\lambda),
\end{equation}
which is applicable to all positive integer $n$. Nevertheless, it can be shown that in fact one can directly analytically continue it to $n=-1$ with the change of the integral region to a circle centered around origin \cite{Boruch:2024kvv}. In this way, we obtain that
\begin{equation}
    (M^{-1})_{ij}=\lim_{n \rightarrow -1}\frac{1}{2 \pi i}\oint d\lambda \frac{1}{\lambda}  R_{ij}(\lambda).
\end{equation}
Now we need to solve the resolvent matrix $\hat{R}(\lambda)$.  We will follow the method presented in \cite{Penington:2019kki}.
As its definition, we have
\begin{equation}
    R_{ij}(\lambda)=(\frac{1}{\lambda-M})_{ij}=\frac{1}{\lambda}\sum_{n=0}^{\infty}\frac{(M^n)_{ij}}{\lambda^n}.
\end{equation}
Each term on the right-handed side of the equation 
corresponds to a product of $n$ inner products.
\begin{equation}
(M^n)_{ij}=\bra{\Psi_i}\ket{\Psi_{l_1}}\bra{\Psi_{l_1}}\dots \ket{\Psi_{l_{n-1}}}\bra{\Psi_{l_{n-1}}}\ket{\Psi_j}.
\end{equation}
Now we can use the results derived in Sec \ref{sec:overlaps} to compute $(M^n)_{ij}$.
We first classify \( (M^n)_{ij} \) by observing that different values of \( n \) correspond to different numbers of boundary circles, as shown in the first equals sign of Fig \ref{fig:resolvent}. Next, for each \( (M^n)_{ij} \), there are various ways to connect them, as demonstrated in Sec \ref{sec:overlaps} and illustrated in the second equals sign of Fig \ref{fig:resolvent}. Furthermore, we find that these diagrams can be reorganized based on the number of connections between the first boundary circle and the subsequent ones. After reorganization, they can be expressed as a geometric series expansion in terms of $\hat{R}(\lambda)$, 
which is shown in the last equal sign of Fig \ref{fig:resolvent}.
Finally, we can get
\begin{equation}
    R_{ij}(\lambda)=\frac{1}{\lambda}\delta_{ij} + \frac{1}{\lambda} \sum_{n=1}^\infty \frac{Z_n}{\lambda^n} R^{n-1}R_{ij}.
\end{equation}
where
\begin{equation}
    R=\sum_{i}R_{ii}(\lambda)
\end{equation}
is the trace. 
In addition, we also need to calculate the matrix elements of the operators $k_L$ and $k_R$ in Eq.(\ref{eq:(3.5)}), for simplicity but without loss of the generality, we consider the following specific operators with the form 
\begin{align}
    k_L &=e^{-\bar{\beta}_{L}H_L},\\
    k_R &=e^{-\bar{\beta}_{R}H_R}.
\end{align}
The action of  such  operators is just adding to the evolution of Hamiltonian, we can easily get  
\begin{equation}
\bra{\Psi_i}k_Lk_R\ket{\Psi_j}=\Tr\left(O^\dagger_i\, \rho_{(\beta_L+\bar{\beta}_L)}\, O_j \, \rho_{(\beta_R+\bar{\beta}_R)}  \right)=\delta_{ij}Z_1((\beta_L+\bar{\beta}_L),(\beta_R+\bar{\beta}_R))\equiv \delta_{ij}Z_1(k_L,k_R).
\end{equation}
Then, similar to the method for solving the recurrence relation for \( R_{ij}(\lambda) \), we can obtain 
\begin{equation}
   R_{ij}(\lambda) \bra{\Psi_j}k_Lk_R\ket{\Psi_i}= \sum_{n=0} ^{\infty}R^{n+1}(\lambda)Z_{n+1}(k_L,k_R).
\end{equation}
In the micro-canonical ensemble, we have
\begin{equation}
    Z(n\beta)=\int dE \rho(E) e^{-n\beta E}=(\Delta E)\, \rho(E)\, e^{-n\beta E},
\end{equation}
and 
\begin{equation}
    \left((\Delta E)\, \rho(E)\right)=e^{S_L+S_R}.
\end{equation}
Therefore, we can obtain 
\begin{align}
 \Tr_{\mathcal{H}_\text{bulk(D)}}(k_L k_R)&= \lim_{n \rightarrow -1}(M^{n})_{ji}\bra{\Psi_i}k_Lk_R\ket{\Psi_j}\\
    &=\lim_{n \rightarrow -1}\frac{1}{2 \pi i}\oint d\lambda \, \lambda^n \sum_{n=0} ^{\infty}R^{n+1}(\lambda)Z_{n+1}(k_L,k_R)\\
    &=\frac{1}{2 \pi i}\oint d\lambda \, \frac{1}{\lambda}\frac{R(\lambda)Z_{n+1}(k_L,k_R)}{1-R(\lambda)e^{-2\beta E}} 
   \end{align}
if $\abs{R(\lambda=0)}=\infty$  (we will prove it in Sec \ref{sec:prove}), then 
\begin{align}
    \frac{1}{2 \pi i}\oint d\lambda \, \frac{1}{\lambda}\frac{R(\lambda)Z_{n+1}(k_L,k_R)}{1-R(\lambda)e^{-2\beta E}}
    &=\frac{\left((\Delta E) \rho(E)\right) e^{(-2\beta E-\bar{\beta}_LE-\bar{\beta}_RE)}}{e^{-2\beta E}} \\
    &=(e^{S_L}e^{-\bar{\beta}_L E})\cdot(e^{S_R}e^{-\bar{\beta}_R E}) \\
    &= \Tr_{\H_L}(k_L)\Tr_{\H_R}(k_R),
\end{align}
Here $\Tr_{\H_L}(k_L)$ represents the boundary trace of $k_L$ in the single-sided bulk Hilbert space. Thus, we have demonstrated the factorisation of the Hilbert space of the two-sided black hole.
In fact, the single-sided Hilbert space with dimension \( e^{S_L} \) corresponding to a single-sided black hole. Moreover, we can obtain more specific information about this single-sided bulk Hilbert space: in the microcanonical ensemble, we can span the single-sided bulk Hilbert space using single-sided thin shell states, as investigated in \cite{Chandra:2022fwi, Balasubramanian:2022lnw, Geng:2024jmm}.
To show it, we can define the trace of $k_L$ in the Hilbert space \( \H_L(d_L) \) formed by thin shell states:
\begin{equation}
    \Tr_{\H_L(d_L)}(k_L)=\left( \bra{\psi_i}\ket{\psi_j}^{-1}\bra{\psi_i}k_L\ket{\psi_j} \right).
\end{equation}
Similar to two-sided black holes, if the  dimension $d_L$ is greater than $e^{S_L}$, we can obtain 
\begin{align}
    \Tr_{\H_L(d_L)}(k_L)&=\lim_{n \rightarrow -1}\frac{1}{2 \pi i}\oint d\lambda \, \lambda^n R_{ij}(\lambda)\bra{\psi_j}k_Lk_R\ket{\psi_i} \\
    &=\lim_{n \rightarrow -1}\frac{1}{2 \pi i}\oint d\lambda \, \lambda^n \sum_{n=0} ^{\infty}R^{n+1}(\lambda)\tilde{Z}_{n+1}(k_L)\\
    &=(e^{S_L}e^{-\bar{\beta}_L E})=\Tr_{\H_L}(k_L).
\end{align}
Therefore, we conclude that a two-sided black hole Hilbert space can be expanded via the tensor product of two single-sided black hole Hilbert space, and the Hilbert space of each single-sided black hole can be spanned by a basis of single-sided thin shell states (although non-orthogonal), namely
\begin{equation}
\label{eq:er=epr-hint}
\H_{\text{bulk(D)}}=\H_{L(d_L)}\otimes\H_{R(d_R)}.
\end{equation}
\subsection{Proving $R(\lambda=0)=\infty$ when $D\ge e^{S_L+S_R}$ }
\label{sec:prove}

In the microcanonical ensemble, we can easily solve for the resolvent matrix:
\[
R(\lambda) = D e^{S_L+S_R} \left( \frac{1}{2} - \frac{y}{2x} - \frac{\sqrt{\frac{y^2}{4} + \frac{x^2}{4} - x \left( 1 + \frac{y}{2} \right)}}{x}  \right)
\]
where
\[
y = \frac{e^{S_L+S_R} - D}{D} \quad \text{and} \quad x = e^{S_L+S_R} \lambda.
\]
From this, we can directly read that when $D\ge e^{S_L+S_R}$, $\abs{R(0)}=\infty$.
In fact, it is easy to see that since the dimension of Gram matrix $\hat{M}$ is D while but its rank is $e^{S_L+S_R}$, when $D> e^{S_L+S_R}$, $M_{ij}$ becomes a degenerate matrix, which must have zero eigenvalues. Therefore, at $ \lambda = 0 $, $R(0)=\Tr(1/M) $ becomes infinite. When $D= e^{S_L+S_R}$ there is no single pole, but the branch cut will hit the point $ \lambda = 0$, we still have $\abs{R(0)}=\infty$.

\section{ From factorisation to ER=EPR}
\label{sec:From factorization to ER=EPR}
Above, we demonstrated that the direct product of two single-sided Hilbert spaces can form a Hilbert space for a two-sided black hole. This brings to mind the ER=EPR conjecture \cite{Maldacena:2013xja}, which suggests that through entanglement, two single-sided black holes can give rise to an Einstein-Rosen bridge (the spatial slice of the two-sided black hole). In this section, we aim to explore the connection between the factorisation of the two-sided Hilbert space and the ER=EPR conjecture. 

Our question is: having established the relationship between the Hilbert spaces of two-sided and single-sided black holes, we want to further explore the correspondence between specific states within these Hilbert spaces. Specifically, how can we form a two-sided black hole state through the superposition of states in the tensor product space of two single-sided black hole Hilbert spaces?  We find that when the superposed states exhibit sufficiently strong entanglement between the two single-sided black holes, a two-sided black hole state can be formed. A realization of such a scenario has appeared in JT gravity \cite{Anderson:2020vwi}. See also \cite{Balasubramanian:2021wgd}\cite{Balasubramanian:2020coy}\cite{Miyata:2021qsm} for studies on the relationship between the entanglement of two eternal black holes and ER=EPR. With the power of thin shell states, we may implement this picture in higher-dimensional gravity.

Consider a superposition state $\ket{\Psi}$ in the tensor product of two single-sided Hilbert spaces 
\begin{equation}
\ket{\Psi}=\sum_{i,j}^{d}c_{ij}\ket{\psi_i}\ket{\psi_j}^*\in \H_{L(d_L)}\otimes\H_{R(d_R)}.
\end{equation}
Here, we consider the microcanonical ensemble, with the left and right black holes having the same ADM mass. Note we apply the CRT conjugate $\ket{\psi_j}^*$ for the purpose of constructing the gravitational dual states. For technical convenience, we impose the following conditions as \cite{Anderson:2020vwi}  on the coefficient matrix
\begin{align}
    c=c^\dagger&=c^2, \\
    \Tr(c)&=d.
\end{align}
Due to the correspondence relationship Eq.(\ref{eq:er=epr-hint}) between the Hilbert spaces,  we know $\ket{\Psi}\in \H_{\text{bulk(D)}}$.
Then we can ask under what conditions does $\ket{\Psi}$ correspond to a connected two-sided black hole, and when does it correspond to two disconnected single-sided black holes? 

We can answer this question by computing the norm of the state $\ket{\Psi}$
\begin{align}
\bra{\Psi}\ket{\Psi}&=\sum^d_{i_1 j_1 i_2 j_2}c_{i_1 j_1}c^*_{i_2 j_2}\bra{\psi_{i_2}}\ket{\psi_{i_1}}\overline{\bra{\psi_{j_2}}\ket{\psi_{j_1}}} \\ &=\sum^d_{i_1 j_1 i_2 j_2}c_{i_1 j_1}c^*_{i_2 j_2}\bra{\psi_{i_2}}\ket{\psi_{i_1}}\bra{\psi_{j_1}}\ket{\psi_{j_2}}.
\end{align}
As we demonstrated in Eq.(\ref{eq:phi square}), the product of two inner products \( \langle \psi_{i_2} | \psi_{i_1} \rangle \langle \psi_{j_1} | \psi_{j_2} \rangle \) will include contributions from both connected saddle $Z_2$ and disconnected saddle $(Z_1)^2$, as illustrated in Fig \ref{eq:phi square}.  Note that the  analytic continuation of the Euclidean saddle point geometry $(Z_1)^2$ will correspond to two separate Lorentzian single-sided thin shell black holes, while $Z_2$ will correspond a Lorentzian connected two-sided black hole geometry. 

The disconnected saddle gives rise to
\begin{equation}
\bra{\Psi}\ket{\Psi}_{dc}=\Tr(cc^{\dagger})Z_1^2\sim \vcenter{ \hbox{\includegraphics[width=0.2\textwidth]{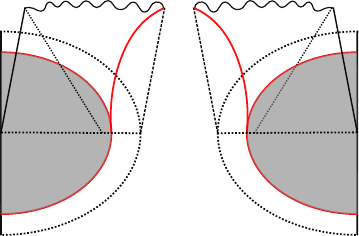}}}\, ,
\end{equation}
and the connected saddle give a two-sided black hole geometry 
\begin{equation}
\bra{\Psi}\ket{\Psi}_{c}=\Tr(c)\Tr(c^{\dagger})Z_2\sim \vcenter{ \hbox{\includegraphics[width=0.2\textwidth]{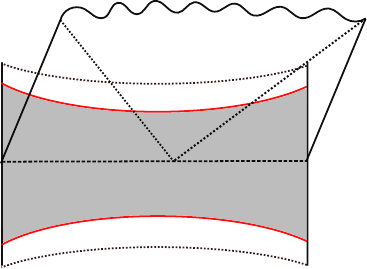}}}.
\end{equation}
Finally, using the microscopic ensemble partition function  we may obtain
\begin{align}
\bra{\Psi}\ket{\Psi}&=\Tr(cc^{\dagger})Z_1^2+\Tr(c)\Tr(c^{\dagger})Z_2 \propto (d e^{2S} +d^2 e^S).
\end{align}

When \( d \gg e^S \), the dominant contribution to the norm \( \bra{\Psi}\ket{\Psi} \) arises from the connected part \( Z_2 \) (wormhole saddle). This can be interpreted as the quantum state having a high probability of corresponding to a connected Lorentzian two-sided black hole (after analytic continuation). Conversely, when \( e^S \gg d \), the dominant contribution comes from the square of the disconnected saddle $Z_1$, whose Lorentzian counterpart corresponds to two disconnected single-sided black holes. Meanwhile, one can use the replica trick to calculate that when \( d \gg e^S \), the entanglement entropy between two  black holes reaches its maximum value, which is equal to the  \( S \). In contrast, if \( e^S \gg d \), the entanglement entropy is   \( \ln(d) \). 

Thus, we conclude that classical connected spacetime geometry emerges when two single-sided black holes are maximally entangled.\footnote{However, our discussion does not address the question of whether the geometry of a typical state is necessarily smooth \cite{Marolf:2013dba}.} Conversely, as the state becomes disentangled, it transitions into two disconnected single-sided black holes.  In our current setup, we have only considered the left side and right side Hilbert space coming from the same set of thin shell states (basis). It implies that the connected geometry solution can be observed only from some special entangled states. On the other hand, how we choose the basis should not affect the state construction, but we did not find a good way to let the two-sided Lorentzian black hole emerge at the level of saddle point solution for general entangled states and general basis, perhaps we need some methods to go beyond the saddle point approximation. This is an interesting problem worth exploring. It would be also interesting to consider the entanglement between two single-sided black holes with different ADM masses and explore how a two-sided black hole geometry emerges between them.

\section{Conclusion and discussion}
In this paper, we examined the factorisation problem of Hilbert spaces in high-dimensional black holes, When considering non-perturbative wormhole effects, we demonstrated that the Hilbert space associated with a two-sided black hole can be factorized into two single-sided Hilbert spaces. 

Our result relies on a key fact: that the infinite-dimensional Hilbert space at the perturbative level becomes effectively finite when considering non-perturbative wormhole effects, as the non-perturbative corrections alter the naive orthogonality of microstates by introducing small but non-zero overlaps. 

Furthermore, we emphasize that the single-sided bulk Hilbert space, after the factorisation of the two-sided black hole, can be spanned by the thin shell states of a single-sided black hole, which evokes the ER=EPR conjecture: The microstates of a two-sided black hole can emerge from the entanglement of two disconnected single-sided black hole states. This reinforces the idea that spacetime connectivity, at least in the context of black holes, can arise from entangled states in tensor product of two Hilbert spaces. We have demonstrated that when the entanglement between two single-sided black holes is sufficiently strong, their inner product is predominantly determined by the connected Euclidean geometry. The Lorentzian continuation of this connected Euclidean geometry corresponds to a two-sided black hole. In contrast, when the entanglement is sufficiently weak, the system consists of two independent single-sided black holes. However, the details of the transition between these two geometric configurations remain unclear and evidently extend beyond the framework of classical geometric descriptions. It may be possible to analyze this transition using an algebraic approach, as suggested in \cite{Engelhardt:2023xer}, within certain models involving bulk quantum fields.

\section*{Acknowledgments}
The author is grateful to Yi Ling for encouragement throughout this work and his improvements to the manuscript. I also thank Kai Li, Yi Ling, Wenbin Pan, and Zhangping Yu for their valuable discussions. Special thanks go to Ronak M. Soni for his insightful correspondence, which helped resolve my confusion.
This work is supported in part by the Natural Science Foundation
of China under Grant No.~12035016 and 12275275. It is also supported by Beijing Natural Science Foundation under Grant No. 1222031, and the Innovative Projects of Science and Technology No. E2545BU210 at IHEP.

\bibliographystyle{JHEP}
\bibliography{biblio.bib}

\end{document}